\documentstyle[prl,aps]{revtex}
\begin{document}
\draft

\twocolumn[\hsize\textwidth\columnwidth\hsize\csname@twocolumnfalse\endcsname

\title{Cracking Piles of Brittle Grains}
\author{Franti\v{s}ek Slanina}
\address{Institute of Physics,
	Academy of Sciences of the Czech Republic,
	Na~Slovance~2, CZ-18221~Praha,
	Czech Republic\\
	and Center for Theoretical Study
	Jilsk\'a~1,CZ-11000~Praha, Czech
	Republic\\
	e-mail: slanina@fzu.cz}
\maketitle

\begin{abstract}
A model which accounts for
cracking 
avalanches in piles of grains subject to external load is introduced
and numerically simulated. The stress is 
stochastically transferred from higher layers to lower ones. Cracked
areas exhibit various morphologies, depending on the degree of
randomness in the packing and on the ductility of the grains. The
external force 
necessary to continue the cracking process is constant in wide range
of values of the fraction of already cracked grains. If the grains
are very brittle, 
 the force fluctuations become periodic in early stages of
cracking. Distribution of cracking avalanches obeys a power law with 
exponent $\tau = 2.4 \pm 0.1$. 
\end{abstract}

\twocolumn]

\section{Introduction}

There are many phenomena concerning granular matter which attract
attention of physicists \cite{ja_na_be_96a}. The source of complexity
of sand and similar    
systems stems from highly non-linear mechanic response on the
mesoscopic scale ({\it i. e. } on the scale of single grains) which
brings about 
complicated behavior on many scales, up to the macroscopic one, even
though there is usually no scale-free behavior \cite{ra_je_mo_ro_96}. 
This feature brings the physics of granular matter close to other
complex mechanics phenomena, like friction \cite{per_to_96} and wear 
\cite{kato_93}, where the
interplay of mesoscopic and macroscopic phenomena is the central point
of attention.

The dynamics of sand may be studied from two points of view. Slow
driving by adding single grains gives rise to avalanches
\cite{nagel_92,he_so_ke_ha_ho_gri_90} and 
stratification phenomena \cite{ma_ha_ki_sta_97}. Intense driving by
periodic or  
persistent external forces was observed to cause for example surface
pattern formation (dunes etc.) or grain-size separation
\cite{ja_na_be_96a}. Dynamics of mixture  of sand and air may lead to
beautiful phenomena like ticking of hour glasses
\cite{lepe_ma_ha_am_bi_wu_96}. 

On the other hand, the most frequently asked question about static
properties was the stress distribution within sand heaps, either
free of embedded in various kinds of containers
\cite{ra_je_mo_ro_96,ou_rou_97,ed_mou_96,mou_ed_96,ed_mou_96a,li_na_sche_co_ma_na_wi_95}. 
The most famous  
phenomenon is perhaps the minimum of stress directly below the top of
a conic sandpile, measured by Smid and Novosad \cite{smi_no_81} and later on
explained theoretically by Bouchaud and others
\cite{bou_ca_cla_95,wi_cla_ca_bou_96,cla_bou_ca_wi_98}. The 
explanation is based on the fact, that arches are created within the
granular packing, which support most of the weight. A very important
phenomena connected with arching are the static avalanches due to
large-scale reconstruction of arches, caused by very small external
perturbation \cite{cla_bou_97}, and stick-slip motion of sand in a tube
\cite{cla_bou_97a,cla_bou_98}. 

Both of the above phenomena are currently well described within the
scalar arching model\cite{cla_bou_97a}, which  is a generalization of
the scalar stress model developed for granular matter by Liu {\it et
al. } \cite{li_na_sche_co_ma_na_wi_95,co_li_ma_na_wi_96,coppersmith_97}. 

Less studied phenomenon from the point of view of granular materials
is the procedure 
in which the grains are produced, {\it i. e. } the fragmentation
process \cite{redner_90,ma_zha_96}. The obvious practical importance
of this process was stressed {\it e. g. } in \cite{bi_gu_og_90}. In
the statistical 
approaches to fragmentation \cite{redner_90}, the 
grains which are cracked are considered either independently of each
other or random two-particle collisions of the grains are taken into
account. Such models are appropriate to the situation in
mills. Different mechanisms should be at work when the bulk of the
heap of granular particles is cracked by compression, like in 
manufacturing pills in pharmaceutical industry. 
Similar problems were already addressed when studying localization of
deformation in two-dimensional heaps of plastic cylinders
\cite{po_am_bi_tro_92} and compaction of granular matter in silos
under pressure \cite{evesque_97a}. 

In the present work, we introduce a model, which considers 
cracking of grains within a pile of other grains, some of them already
cracked, others not. So, we will not investigate the size distribution
of fragments, like in Ref. \cite{redner_90}, but the spatial configuration of
clusters of cracked grains and also the external force fluctuations occurring during the process of cracking.

The article is organized as follows. In the next section the model is
introduced. The section III is a gallery of simulation results and the
last section, sect. IV draws conclusions from the results obtained.

\section{Description of the model}
Our model describes a two-dimensional pile of granular matter contained
in a rectangular silo. A physical realization of this situation may be
prepared by two parallel glass plates, distance of which corresponds
to grain size. The lateral and bottom slots are closed, while the
upper slot is open and a uniform external force is applied to the
surface of the pile by a kind of piston.  The grains are brittle (eggs
may serve as a popular example), which means that if the stress the
grain supports
exceeds a threshold value $w_{\rm thr}$, the grain collapses. As a
consequence of this, the stress pattern in the neighborhood of the
collapsed grain changes, which may cause another grain collapse and
finally leads to a kind of internal avalanche. During that process, the
piston is kept immobile, so the total external force decreases, until the
avalanche stops. How much the force decreases as a consequence of
cracking one grain, is described by a material dependent factor
$\alpha < 1$. 
We may expect, that for more ductile grains, the drop of the force
will be smaller and the parameter $\alpha$ will be closer to 1. For
this reason we will call $\alpha$ the ductility.

The stress within the pile is a tensor, but recent studies
\cite{cla_bou_ca_wi_98} 
showed that for many purposes only the diagonal element corresponding
to the horizontal axis is important. This simplification leads to a
scalar model of stress propagation in granular matter, which will be a
basis of our model here.

We suppose the grains are placed regularly on a square lattice rotated
by 45 degrees, so that the columns and rows of grains correspond to
the diagonals on the lattice. Each row is $L$ grains wide, each column
is $H$ grains high. The grains are in contact with the nearest
neighbors on the lattice. The randomness in the size, shape, and position of
the grains is taken into account by a stochastic rule, which describes the
propagation of stress. 

Denote $w_{ik}$ the stress on the grain in $i$-th row (counted from
above) and $k$-th column. It transfers the fraction $q_{ik}$ of the
stress to its left bottom neighbor, the fraction $1-q_{ik}$ to its
right bottom neighbor. We neglect the weight of the grains themselves,
compared to the external force. So, the rule of stress propagation is
described by the equations
\begin{equation}\left.\begin{array}{lll}
w_{i+1,k}&=q_{ik}w_{ik} + (1-q_{ik-1})w_{ik-1}&\; 
{\rm for\; odd\; {\it k} }\\
w_{i+1,k}&=(1-q_{ik})w_{ik} + q_{ik+1}w_{ik+1}&\; 
{\rm for\; even\; {\it k} }\; .
\label{eq:q-rule}
\end{array}\right.\end{equation}
We impose cylindrical boundary conditions, $w_{i0} = w_{iL}$.
The topmost row is subject to external force
$w_{1k} = f_k$. Total force on the piston is then $F=\sum_k f_k$.

The simulation proceeds as follows. 
The numbers $q_{ik}$ are taken randomly from the uniform distribution
on the interval $({1-\beta\over 2},{1+\beta\over 2})$. 
Initially all $f_k$ are set equal and the local stresses are computed
according to rules (\ref{eq:q-rule}). The force is increased until stress on
one non-cracked grain, say at position $(i,k)$, reaches the threshold
$w_{\rm thr} = 1$. Then, the grain is cracked, which has two
consequences. First, the 
force on top of its column is lowered, $f_k\to \alpha f_k$. Then, if
grain in the same row to the left, {\it i e. } $(i,k-1)$ is not
cracked, the value of $q$ corresponding to 
left top neighbor of $(i,k)$ is set to 1. If $(i,k-1)$ is cracked $q$
is given new random value from the uniform distribution on the interval
$({1-\beta\over 2},{1+\beta\over 2})$. Similar rule applies on
the right hand side: if $(i,k+1)$ is not cracked, the right top
neighbor if $(i,k)$ has new $q=0$, if $(i,k+1)$ is cracked, the new
$q$ is a random number from the same distribution as above. These
rules correspond to very simple intuitive observation, that the
cracked grain does no more bear the load, if it has neighbors, which
can bear the load instead of it. However, if the neighbors are also
already cracked grains, the stress propagation remains to be
stochastic as it was before the cracking, but the realization of the
randomness, {\it i. e. } the values of the numbers $q$ are changed.

After each change of $q$'s, the local stresses are recomputed, the
grains which are not yet cracked and exceed the threshold are cracked,
new $q$'s are established and this procedure is repeated until no
non-cracked grains exceeding the threshold are found. Then, the external
force is increased up to the value when another grain is cracked
again and new cracking avalanche begins. We will call avalanche size
$s$ total number of grains cracked during the avalanche. 
This algorithm continues as long as there are any non-cracked grains
left.

Besides the size of the system, the model has two free parameters. The
parameter $\alpha$ measures the ductility of the grains and $\beta$
the degree of randomness in the stress propagation. The limit case
$\beta=0$ corresponds to fully deterministic case.

\section{Simulation results}

\begin{figure}[hb]
  \centering
  \vspace*{80mm}
  \includegraphics{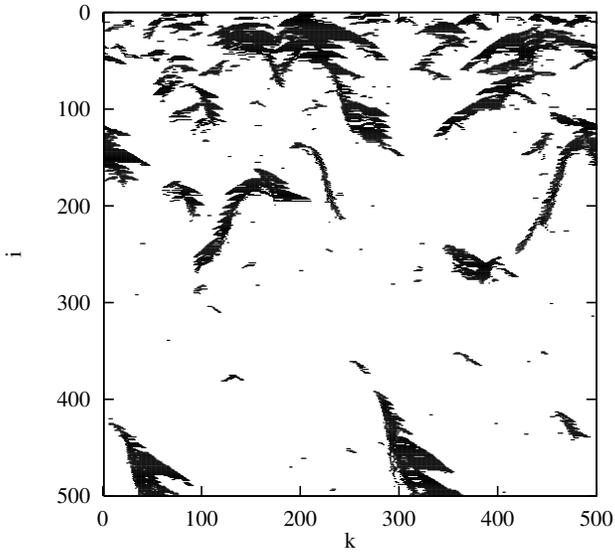}
  \caption{Morphology of cracked areas for a sample with $L=500$, $H= 500$,
  after 5000 time steps. Every cracked grain is represented by a
  black dot. The   parameters are $\alpha = 0.9$, $\beta= 0.25$.}
  \label{fig:morph-9-25}
\end{figure}

\begin{figure}[hb]
  \centering
  \vspace*{80mm}
  \includegraphics{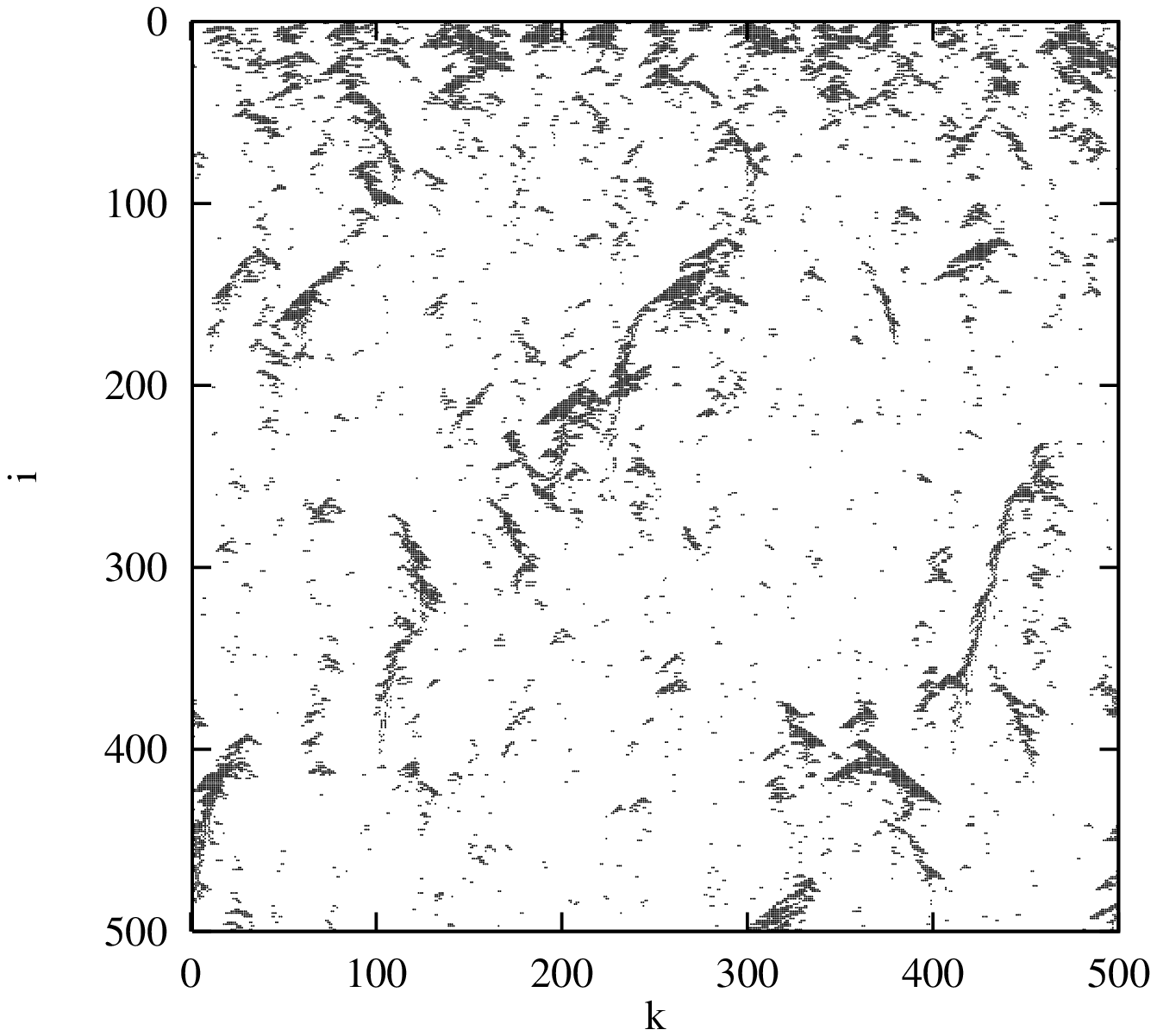}
  \caption{Morphology of cracked areas for a sample with $L=500$, $H= 500$,
  after 5000 time steps. The   parameters are $\alpha = 0.9$,
  $\beta= 0.5$.} 
  \label{fig:morph-9-5}
\end{figure}

\begin{figure}[hb]
  \centering
  \vspace*{80mm}
  \includegraphics{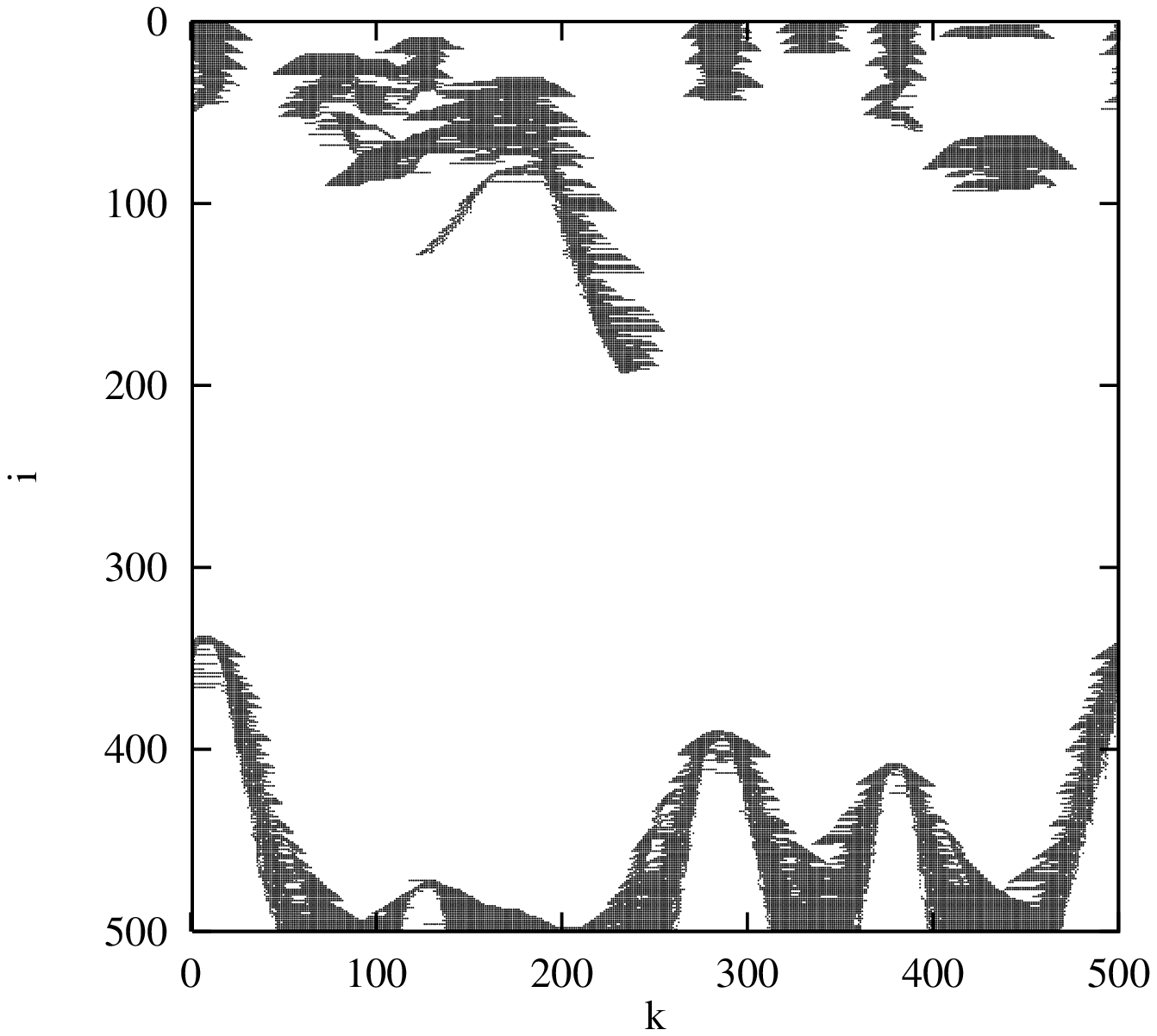}
  \caption{Morphology of cracked areas for a sample with $L=500$, $H= 500$,
  after 5000 time steps. The   parameters are $\alpha = 0.9$,
  $\beta= 0.1$.}  
  \label{fig:morph-9-1}
\end{figure}

\begin{figure}[hb]
  \centering
  \vspace*{80mm}
  \includegraphics{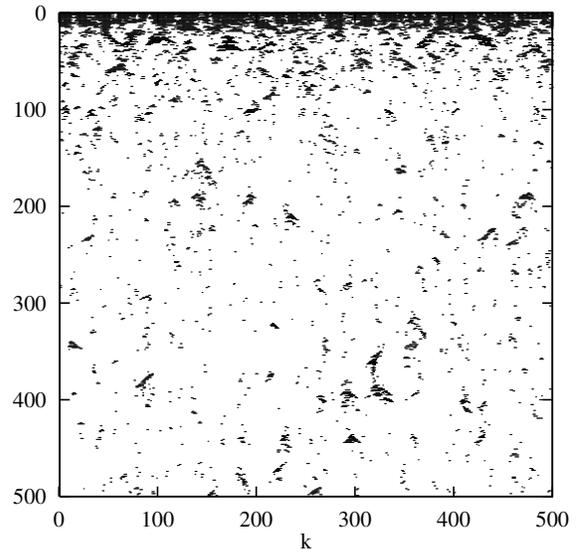}
  \caption{Morphology of cracked areas for a sample with $L=500$, $H= 500$,
  after 10000 time steps.
  The   parameters are $\alpha = 0.5$, $\beta= 0.5$.}
  \label{fig:morph-5-5}
\end{figure}

When a grain is cracked, the load is mostly transferred to its
neighbors, which have then increased chance to be cracked. This leads
to creation of clusters of cracked grains, which grow and merge as the
cracking proceeds. Typical morphology of the cracked clusters is shown
in the Fig. \ref{fig:morph-9-25}. We can observe formation of
``arches'' with one dominant ``leg'' only. The shape of the ``legs''
resembles the letter S when they grow large. The dependence of the
morphology on the ductility $\alpha$ and randomness $\beta$ is
shown in the Figs. \ref{fig:morph-9-5}, \ref{fig:morph-9-1}, and
\ref{fig:morph-5-5}.  For larger $\beta$ the typical size of the
cracked clusters is smaller, while for small $\beta$ the sample
contains only few big ``arches'', which are also more symmetric than
those for larger randomness. The ductility has different influence
on the morphology: in the case of more brittle grains, {\it i. e. } with
smaller 
$\alpha$, the cracked areas are mostly concentrated in the top part of
the sample, while more ductile grains lead to cracking equally
probable in the whole bulk of the sample. (We performed simulations
also for very ductile grains, $\alpha$ close to 1, and the trend was
observed to shift the cracked regions to the bottom of the sample,
when the ductility is increased.) 


\pagebreak

\begin{figure}[hb]
  \centering
  \vspace*{80mm}
  \includegraphics{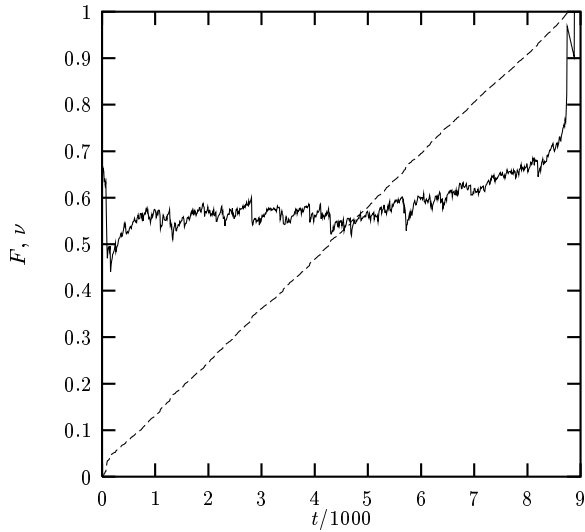}
  \caption{Time evolution of external force $F$ (full line) and
  fraction of cracked 
  grains $\nu$ (dashed line) for the sample with $L=200$, $H=200$,
  $\alpha=0.9$, 
  $\beta=0.25$.}
  \label{fig:force-9-25}
\end{figure}

When the cracking proceeds, the force necessary to continue
fluctuates. Each cracking avalanche means a drop of the force, which
then rises again. The Fig. \ref{fig:force-9-25} shows the time
dependence of the external force $F$ and the fraction of cracked
grains $\nu$ for the sample of $200\times 200$ grains. 
We can see that
the force fluctuates around a nearly time-independent value $F_{\rm
av}\simeq 0.55$ during large part of the process, at least from time
$t=1000$ to $t=5000$. This was even more clearly observed for larger
samples (in our simulations $500\times 500$). So, the picture of the
overall behavior of the force can be as follows. After a transient
period, where the force suddenly drops and slowly rises again, a
stationary cracking regime develops, %
characterized by constant average force
$F_{\rm av}$. This regime holds if the fraction of cracked grains is
small, according to our observations $\nu < \nu_{\rm max}$
is sufficient condition, where the value of $\nu_{\rm max}$ depends
slightly on $\alpha$. For $\alpha=0.1$ we found  $\nu_{\rm max} \simeq
0.7$, while for $\alpha=0.9$ we observed $\nu_{\rm max} \simeq 0.4$.

The value of stationary force $F_{\rm av}$ decreases with $\beta$. we
found the values in the range from $F_{\rm av}\simeq 0.3$ for $\beta=1$
(maximum randomness) to $F_{\rm av}\simeq 0.6$ for $\beta=0.1$
(minimum randomness studied).

Around the average force, there are fluctuations, which reflect unique
realization of the disorder in our sample. We investigated statistical
properties of the fluctuations, using histogram of the changes of
force from one time step to the next one. The distribution of upward
changes can be very well fitted by an exponential, while the downward
changes do not have any clear form of distribution: neither Gaussian,
exponential, stretched exponential nor power-law fit was satisfactory.
A distribution with a power-law tail seems to be a good
candidate, but further data would be needed to settle this question.

\begin{figure}[hb]
  \centering
  \vspace*{80mm}
  \includegraphics{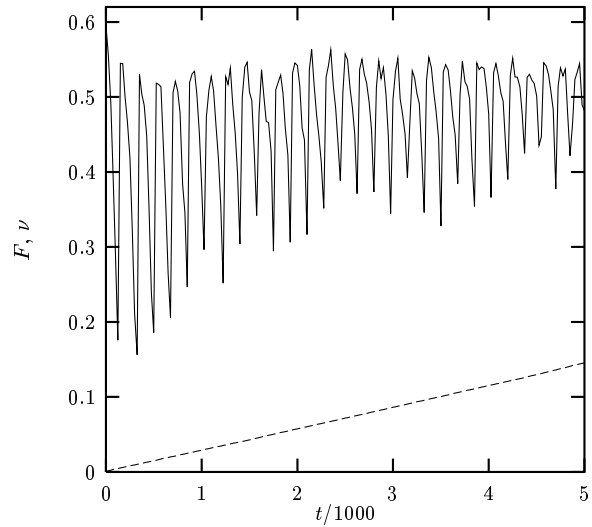}
  \caption{Time evolution of external force $F$ (full line) and
  fraction of cracked 
  grains $\nu$ (dashed line) for the sample with $L=200$, $H=200$,
  $\alpha=0.1$,   $\beta=0.25$.}
  \label{fig:force-1-25}
\end{figure}

For very small $\alpha$ (we observed the phenomenon for $\alpha=0.1$,
but for $\alpha=0.3$ it was already absent) the fluctuations loose
their purely random appearance and quasi-regular force oscillations
occur, which are especially pronounced in the early stages of the
cracking process ({\it i. e. } for small $\nu$). They can be clearly
seen in the Fig. \ref{fig:force-1-25}. When the fraction of cracked
grains increases, the oscillations gradually disappear. The
oscillations perhaps correspond to the sudden drop of the force,
observed for all $\alpha$, followed by gradual increase of the force
again. While for small $\alpha$ many periods of the oscillation may be
realized, for larger $\alpha$ the oscillations are ``over-damped'' and
only single period occurs.

A cracking avalanche starts from the stable state, in which stresses
on all non-cracked grains is below the threshold. The avalanche is
initiated by increase of external force up to value which causes one
grain to crack. This cracking may result in cracking other grains, and
so on, until new stable state is reached and the avalanche stops. We
denote $\Delta c$ the avalanche size, which is the number of grains
cracked during the avalanche. We are interested in statistical
distribution of avalanche sizes. We expect, that the distribution may
be different in the initial transient period and in the stationary
regime, in which the average force $F_{\rm av}$ is constant. So, we
investigated the distributions $P_l^>(\Delta c)$ defined as
probabilities that the size of the avalanche, occurring in time interval
$(t_{l-1},t_l]$, with $t_0=0$, is larger than $\Delta c$.

\pagebreak

\begin{figure}[hb]
  \centering
  \vspace*{80mm}
  \includegraphics{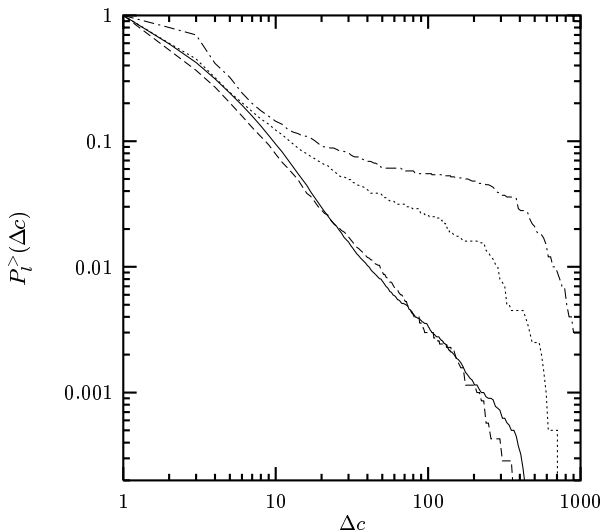}
  \caption{Avalanche size distributions for $L=500$, $H=500$,
  $\alpha=0.9$, $\beta=0.25$, in intervals determined by times
  $t_1=100$, $t_2=300$, $t_3=1000$, and $t_4=5000$. The lines denote
  the following distributions: dash-dotted line $P^>_1$, dotted line
  $P^>_2$, dashed line $P^>_3$, solid line $P^>_4$. $P_l^>$
  corresponds to the interval $(t_{l-1},t_l]$.}
  \label{fig:aval-9-25}
\end{figure}

Figure \ref{fig:aval-9-25} shows the results for a $500\times 500$
sample. The first two intervals, with final times $t_1 = 100$ and $t_2
= 300$ describe 
the situation in the transition regime. We can see that most of
avalanches have typical size about $\Delta c \simeq 400$. On the other
hand, the next two intervals with end times $t_3 = 1000$ and $t_4 =
5000$ give distributions which can be fitted by a power law in the
range  of two decades. It can be also seen that the distribution is
stable in time during the stationary cracking regime. We fitted the
exponent of the power-law dependence $P^>(\Delta c)\sim (\Delta c
)^{1-\tau}$ with the result
\begin{equation}
\tau = 2.4 \pm 0.1 \;\; .
\end{equation}

We have found the same exponent (within error bars) for all values of
parameters studied. The only exception was the case of $\beta = 1$,
where the distribution was close to exponential, instead of power-law.
The breakdown of  power-law, when $\beta$ approaches 1 remains to be
studied. 

\section{Conclusions}

We have found that the two-dimensional pile of brittle grains packed
in a rectangular container exhibit nontrivial behavior, when an
external force is applied from above and the grains are cracked. The
cracked grains form clusters with different morphologies, depending on
the ductility of the grains and on the degree of randomness in the
packing. Degree of randomness seems only to determine the
characteristic scale of the cracked clusters: lower randomness leads
to larger clusters. This fact can be understand rather easily, if we
realize that a cluster occurs when the local stress exceeds the
threshold necessary for a grain to be cracked. If the stress
distribution is more uniform, fluctuations above the threshold are
more distant one from the other.

Less expected feature is the influence of the ductility. Brittle
grains have 
tendency to crack in the top part of the container, while ductile
grains are cracked mostly in the bottom part. This finding may play
important role in separation of grains of different types.

During the cracking process the external force fluctuates around a
general trend, which can be described as follows. If the grains are
not too brittle ($\alpha \gtrsim 0.3$), the force drops suddenly and
then rises slowly to a value, which then remains constant for great
part of the whole cracking process. When the fraction of cracked
grains approaches 1, the force increases again. So, there exists
well-defined stationary cracking regime, preceded by a transient
period and followed by a final stage. For very brittle
grains, the force oscillates rather regularly even in the
stationary regime.

Cracking one grain may result in an avalanche of further
crackings. The distribution of avalanche sizes depends on time. While
in the transient period the distribution is not scale-invariant, in
the stationary regime the distribution of avalanche sizes obeys a power
law. This is an indication, that a sort of criticality is present in
the cracking process. The value of the exponent $\tau\simeq 2.4$ is
larger than avalanche exponents found in most self-organized critical
models known to 
us. On the
other hand, the dynamics of our model resembles the
Olami-Feder-Christensen model of earthquakes \cite{ola_fe_chri_92},
where the exponent varies in wide range, comprising also the value
found in our model. However, the mechanism leading to power-law
scaling in the OFC model is not completely clear. 
This may suggest that a new mechanism leading to criticality is at
work here, different from the usual SOC. 

This work does not compare the simulation results with experimental
data, because we were not able to find any report of an experiment of
this kind (loosely related are the experiments reported in
\cite{po_am_bi_tro_92}). It would be very welcome if a measurement in
the direction suggested here was done in future.

\acknowledgments{
I wish to thank E. Guyon for useful discussions and B. Velick\'y for
inspiring comments which motivated this work.
}

\end{document}